\documentclass[showpacs,prl,preprint,superscriptaddress]{revtex4}

\usepackage[latin1]{inputenc}
\usepackage{graphics}
\usepackage{dcolumn}
\usepackage{bm}

\newcommand {\nc} {\newcommand}
\nc {\beq} {\begin{eqnarray}}
\nc {\eeq} {\end{eqnarray}}
\nc {\eeqn} [1] {\label{#1} \end{eqnarray}}
\nc {\eol} {\nonumber \\}
\nc {\eoln} [1] {\label{#1} \\}

\begin{document}

\title{Multichannel coupling with supersymmetric quantum mechanics \\
and exactly-solvable model for Feshbach resonance}
\author{Jean-Marc Sparenberg}
\affiliation{Physique Quantique, C.P.\ 229, Université Libre de Bruxelles, B 1050 Bruxelles, Belgium}
\author{Boris F.\ Samsonov}
\affiliation{Physics Department, Tomsk State University, 36 Lenin Avenue,
634050 Tomsk, Russia}
\author{François Foucart}
\affiliation{Physique Quantique, C.P.\ 229, Université Libre de Bruxelles, B 1050 Bruxelles, Belgium}
\author{Daniel Baye}
\affiliation{Physique Quantique, C.P.\ 229, Université Libre de Bruxelles, B 1050 Bruxelles, Belgium}
\date{\today}

\begin{abstract}
A new type of supersymmetric transformations of the coupled-channel radial Schrödinger equation is introduced,
which do not conserve the vanishing behavior of solutions at the origin.
Contrary to usual transformations,
these ``non-conservative'' transformations allow, in the presence of thresholds,
the construction of potentials with coupled scattering matrices from uncoupled potentials.
As an example,
an exactly-solvable potential matrix is obtained which provides a very simple model
of Feshbach-resonance phenomenon.
\end{abstract}


\maketitle

During the past twenty years,
Darboux transformations of the one-dimensional Schrödinger equation,
also known as supersymmetric quantum mechanics \cite{valladolid:04},
have evolved into a powerful tool
to solve the inverse scattering problem for a given partial wave \cite{chadan:89,baye:04}.
Indeed, this formalism allows both the deductive construction from scattering data
of a unique interaction potential without bound state \cite{sparenberg:97a}
and the construction of all potentials,
phase-equivalent to that potential but displaying different bound spectra
\cite{baye:87a}.
When the initial potential is zero,
the transformed potentials not only fit experimental data with high precision;
they are also exactly solvable and have compact analytical expressions \cite{samsonov:02},
a striking feature of the method
which reveals its efficiency.

This success seems however restricted to the single-channel case:
for coupled channels, supersymmetric transformations were not able up to now to provide a complete
solution of the scattering inverse problem,
despite the early generalization of their algebraic formalism to coupled equations \cite{amado:88a}.
Whereas the construction of phase-equivalent potentials has proved to be possible \cite{sparenberg:97b},
a deductive construction of potentials by inversion of coupled-channel scattering
matrices is still missing.
In the case of coupled channels with different thresholds,
such an inversion even seems impossible, as a matter of principle:
applying a supersymmetric transformation to a diagonal (and hence uncoupled) potential matrix
may result in a coupled potential matrix \cite{chabanov:99} but
the corresponding scattering matrix will always remain diagonal \cite{cannata:93}.
This makes it impossible to invert realistic data with coupling,
which correspond to non-diagonal scattering matrices,
by supersymmetric transformations of simple uncoupled potentials like the zero potential.

The need for coupled-channel inversion techniques is however strongly growing
these days since many problems in physics rely on coupled-channel systems.
In nuclear physics for instance,
theoretical calculations of hypernuclei \cite{nemura:05} require baryon-baryon interactions,
which naturally display coupling between channels corresponding to different hyperon-nucleon
and hyperon-hyperon states.
In the context of experimental Bose-Einstein condensation,
atom-atom interactions are monitored with the help of magnetic Feshbach resonances \cite{feshbach:92},
a phenomenon based on the coupling between different spin states
displaying distinct thresholds in the presence of an external magnetic field \cite{tiesinga:93}.

The purpose of this Letter is to show that the limitations on coupled-channel
supersymmetric transformations mentioned in Ref.\ \cite{cannata:93} do not hold
for a category of transformations not used up to now.
These transformations do not conserve boundary behaviors of solutions of the Schrödinger equation
at the origin,
which is why we call them ``non-conservative''.
We establish general properties of these new transformations,
compare them with usual transformations and finally apply them in the simplest possible situation:
a single transformation of an initial vanishing potential.
This leads to a Feshbach-resonance exactly-solvable model.

Using definitions from coupled-channel scattering theory \cite{taylor:72,newton:82},
we consider a set of $N$ coupled radial Schrö\-din\-ger equations,
that read in reduced units
\beq
-\psi''(k,r) + V(r) \psi(k,r) = k^2 \psi(k,r),
\eeqn{schr}
where prime means derivation with respect to the distance $r$ between the two bodies.
Potential $V$ is an $N \times N$ real symmetric matrix,
supposed to be short ranged and bounded.
The complex wave-number diagonal matrix $k$ is defined by its elements $k_i=\sqrt{E-\Delta_i}$,
where $E$ is the center-of-mass energy and
$0=\Delta_1\le\Delta_2\le\dots\le\Delta_N$ are the threshold energies of the $N$ channels.
The solution $\psi$ is an $N \times N$ solution matrix made of $N$ solution vectors.

Equation (\ref{schr}) being of second order,
its general solution can be constructed by linear combination,
with matrix multiplication on the right-hand side,
of two solution matrices made of $2N$ linearly-independent vectors.
In the following, we use two such pairs of solutions:
Jost solution matrices $f(\pm k,r)$, that behave asymptotically as pure exponentials,
\beq
f(\pm k,r) \mathop{\sim}_{r\rightarrow\infty} \exp(\pm \imath k r),
\eeq
and a pair of solutions made of the regular solution matrix $\varphi(k,r)$,
which vanishes at the origin,
and solution matrix $\eta(k,r)$, the derivative of which vanishes at the origin.
We normalize these as $\varphi'(k,0)=\eta(k,0)=I$,
where $I$ is the identity matrix.
In terms of Jost solutions, the regular solution reads \cite{newton:82}
\beq
\varphi(k,r)=\frac{1}{2\imath}\left[f(k,r)k^{-1}F(-k)-f(-k,r)k^{-1}F(k)\right],
\eeqn{phif}
which defines the Jost matrix $F(k)$ as
\beq
F(k)=f^T(k,0),
\eeqn{Fdef}
where $T$ means transposition.
This is obtained by calculating, both at the origin and at infinity,
the Wronskian
$W[\varphi(k,r),f(k,r)] \equiv \varphi^T(k,r) f'(k,r) - \varphi'^T(k,r) f(k,r)$,
the value of which is independent of $r$,
taking into account $W[f(-k,r),f(k,r)]=2\imath k$.
The Jost matrix is a key quantity in scattering theory:
bound (resp.\ resonant) states correspond to zeros of its determinant in the upper (resp.\ lower)
$k_i$ planes
and the scattering matrix $S$ reads
\beq
S(k)=k^{-1/2} F(-k) F^{-1}(k) k^{1/2}.
\eeqn{S}
For that reason, the following study mostly concentrates on Jost-matrix properties.

To perform a supersymmetric transformation of Eq.\ (\ref{schr}),
we follow Ref.\ \cite{amado:88a} and first factorize it as
\beq
\left[ A^+ A^- - \kappa^2 \right] \psi(k,r) & = & k^2 \psi(k,r),
\eeqn{fact}
where $A^\pm = \pm d/dr + U(r)$ are mutually-adjoint first-order differential operators,
defined in terms of superpotential real symmetric matrix $U$.
Equation (\ref{fact}) is equivalent to Eq.\ (\ref{schr}) when
\beq
U(r)=\sigma'(r) \sigma^{-1}(r),
\eeqn{U}
where $\sigma(r)$ is a solution matrix of Eq.\ (\ref{schr})
at negative energy $\cal{E}$,
referred to as the factorization solution.
This solution matrix is assumed to be real and invertible for all $r$;
moreover, its self-Wronskian $W(\sigma,\sigma)$ vanishes
in order for $U$ to be symmetric.
Another important property of Eq.\ (\ref{U}) is that multiplying the factorization solution
on the right by an arbitrary constant regular matrix,
which is equivalent to changing the factorization solution,
does not affect the superpotential.
This strongly reduces the number of possible transformations.
Finally, the wave-number diagonal matrix $\kappa$ is defined
by its positive elements $\kappa_i=\sqrt{\Delta_i-{\cal E}}$.

We then apply $A^-$ on the left to Eq.\ (\ref{fact}),
which leads to a new Schrödinger equation with potential matrix
\beq
\tilde{V}(r)=V(r)-2 U'(r)=-V(r)-2\kappa^2+ 2 U^2(r)
\eeqn{Vtilde}
and solutions
\beq
\tilde{\psi}(k,r)=A^- \psi(k,r).
\eeqn{psitilde}
The asymptotic behavior of this relation when $\psi$ is a Jost solution
of the initial equation leads to the following expression for the Jost solution
of the new equation
\beq
\tilde{f}(k,r)=A^- f(k,r) [U(\infty)-\imath k]^{-1}.
\eeqn{ftilde}
As will be proved elsewhere,
when all thresholds are distinct,
$U(\infty)$ is a diagonal matrix with elements $\pm\kappa_i$,
the signs depending on the asymptotic behavior of factorization solution $\sigma$.
Let us stress that Eq.\ (\ref{ftilde}) shows that
supersymmetric transformations always transform a Jost solution into a Jost solution,
up to a multiplicative factor;
in particular,
when $f$ vanishes at infinity, $\tilde{f}$ vanishes too.

The situation is very different for the behavior at the origin:
when applied to a regular solution of the initial equation,
Eq.\ (\ref{psitilde}) does not systematically lead to a regular solution of the new equation,
which leads to the distinction between two types of transformations.
We call ``conservative'' transformations those that transform
a regular solution of the initial equation into a regular solution of the new equation:
$\tilde{\varphi}(k,r)\propto A^- \varphi(k,r)$.
Most transformations used up to now in the literature \cite{amado:88a,sparenberg:97b}
belong to this category.
Using Eqs.\ (\ref{phif}) and (\ref{Fdef}),
these transformations can be shown to lead to simple modifications of the Jost matrix,
similar to the single-channel case \cite{sukumar:85c,sparenberg:97a},
\beq
\tilde{F}(k)= [\mp U(\infty) - \imath k]^{\pm 1} F(k).
\eeqn{Fmodcons}
In this equation, the upper (resp.\ lower) signs correspond to $\sigma$
diverging (resp.\ vanishing) at the origin,
two cases which will be considered in detail elsewhere.
For vanishing $\sigma(0)$, Eq.\ (\ref{Fmodcons}), combined with Eq.\ (\ref{S}),
leads for instance to the scattering-matrix modification by supersymmetric transformations
obtained in Ref.\ \cite{amado:88a}.

For one-channel elastic scattering,
Eq.\ (\ref{Fmodcons}) is most useful in the context of inverse scattering problem:
such transformations can easily be iterated,
which leads to a Padé approximant of the Jost matrix of arbitrary order,
used to fit experimental scattering data \cite{baye:04}.
In the coupled-channel case with distinct thresholds, however,
this Jost-matrix modification is very restrictive:
since both $k$ and $U(\infty)$ are diagonal matrices,
an initial decoupled potential always transforms into a potential with a decoupled Jost matrix,
and hence a decoupled scattering matrix,
as found in Ref.\ \cite{cannata:93}.
Consequently, these transformations are not able to fit data corresponding to
non diagonal scattering matrices.

In the present work, we consider a new category of supersymmetric transformations:
those for which the factorization solution is a finite regular matrix at the origin.
In this case, the value of the superpotential at the origin, $U(0)$,
is also finite and can be fixed arbitrarily.
This can be shown by expressing $\sigma$ as a linear combination of $\varphi$ and $\eta$
satisfying Eq.\ (\ref{U}):
\beq
\sigma(r)= \eta(\imath \kappa, r) + \varphi(\imath \kappa, r) U(0),
\eeq
where the assumption $\sigma(0)=I$ does not reduce the generality of the superpotential.
Applying Eq.\ (\ref{psitilde}) to a regular solution of the initial equation
leads in this case to a solution of the new equation which does not vanish at the origin.
This type of supersymmetric transformations hence break boundary conditions
and we propose to call them ``non-conservative''.
Similar transformations have already been used in Ref.\ \cite{samsonov:03}
but for the very particular case of zero factorization energy in the single-channel case.
Using Eqs.\ (\ref{Fdef}) and (\ref{ftilde}),
one gets the modification of the Jost matrix for non-conservative transformations
\beq
\tilde{F}(k)= [U(\infty) - \imath k]^{-1} \left[ F(k) U(0) - {f'}^T(k,0) \right].
\eeqn{Fmod}
This relation is clearly more complicated than for conservative transformations,
with the drawback that its iteration might be difficult.
However, it presents the enormous advantage that even when the initial potential
is uncoupled, which means its Jost solution and matrix are diagonal,
the transformed Jost matrix becomes coupled by choosing a non-diagonal $U(0)$.
This new type of supersymmetric transformation thus seems to solve the main
drawback of supersymmetric quantum mechanics for coupled-channel systems
with distinct thresholds stressed in Ref.\ \cite{cannata:93}.

Let us now illustrate these findings by applying the above formalism
to an initial vanishing potential $V=0$,
for which $F(k)=S(k)=I$.
The most general factorization solution matrix for a non-conservative transformation
can be written in two equivalent ways,
\beq
\sigma(r) & = & \cosh(\kappa r) + \sinh(\kappa r) \kappa^{-1} U(0) \eol
& = & \exp(\kappa r) C + \exp(-\kappa r) D.
\eeqn{sigma}
For $N$ channels, the transformed potential,
as obtained from Eqs.\ (\ref{U}-\ref{Vtilde}),
depends on $\kappa_1, \dots, \kappa_N$
(or equivalently on $N-1$ thresholds and one factorization energy)
and on $N(N+1)/2$ arbitrary parameters appearing in $U(0)$.
It is exactly solvable and is equivalent to the potential
derived by Cox \cite{cox:64,hesse:98} in his $q=1$ case.
However, the present derivation
not only leads to a much simpler analytical expression for the potential
[compare Eq.\ (4.7) of Ref.\ \cite{cox:64} to our Eqs.\ (\ref{U}) and (\ref{Vtilde})]
but also subsumes it:
the restriction $\det A\ne 0$ of Ref.\ \cite{cox:64},
which is equivalent to $\det C\ne0$ in our Eq.\ (\ref{sigma}),
does not apply here.

In the case $\det C\ne0$, one has $U(\infty)=\kappa$,
as shown by Eqs.\ (\ref{U}) and (\ref{sigma}).
According to Eq.\ (\ref{Fmod}),
the Jost matrix of the new potential then reads
\beq
\tilde{F}(k)=(\kappa-\imath k)^{-1}\left[ U(0)-\imath k \right].
\eeqn{Fcox}
For $N=2$ channels for instance,
the potential depends on five real parameters
\beq
\kappa=\left(
\begin{array}{cc} \kappa_1 & 0 \\ 0 & \kappa_2 \end{array}
\right),
\quad
U(0)=\left(
\begin{array}{cc} \alpha_1 & \beta \\ \beta & \alpha_2 \end{array}
\right)
\eeqn{param}
and Eq.\ (\ref{Fcox}) is equivalent to Eq.\ (5.1) of Ref.\
\cite{cox:64}.

As stressed above, our formalism is also valid for $\det C=0$.
In this case, one has $U(\infty)={\rm diag}(-\kappa_1,\kappa_2)$,
which leads to the Jost matrix
\beq
\tilde{F}(k)=\left(
\begin{array}{cc} \frac{k_1+\imath \alpha_1}{k_1-\imath \kappa_1} &
\frac{\imath \beta}{k_1-\imath \kappa_1} \\ \frac{\imath \beta}{k_2+\imath \kappa_2} & 
\frac{k_2+\imath \alpha_2}{k_2+\imath \kappa_2} \end{array}
\right).
\eeqn{Fgcox}
This new result cannot be obtained from Cox' formula.
Moreover, for $\det C=\det D=0$,
which is equivalent to the particular choice of parameters
\beq
\alpha_{1,2}=\pm\sqrt{(\kappa_1 \kappa_2-\beta^2)(\kappa_1/\kappa_2)^{\pm 1}}
\quad (\kappa_1 \kappa_2 > \beta^2),
\eeqn{alpha12}
further simplifications occur.
One gets superpotential
\beq
U(r) = 
\frac{1}{\cosh y}
\left( \begin{array}{cc}
-\kappa_1  \sinh y& \sqrt{\kappa_1 \kappa_2} \\ 
\sqrt{\kappa_1 \kappa_2} & \kappa_2 \sinh y
\end{array} \right)
\eeq
with $y = (\kappa_2 - \kappa_1) r - {\rm arccosh} \sqrt{\kappa_1 \kappa_2/\beta^2}$
and potential
\beq
{\tilde V} = 
\frac{2(\kappa_2 - \kappa_1)}{\cosh^2 y}
\left( \begin{array}{cc}
\kappa_1 & \sqrt{\kappa_1 \kappa_2} \sinh y \\ 
\sqrt{\kappa_1 \kappa_2} \sinh y & -\kappa_2 
\end{array} \right)
\eeqn{Vgcox}
where diagonal term $\tilde{V}_{11}$ (resp.\ $\tilde{V}_{22}$) is positive (resp.\ negative).
The determinant of Jost matrix (\ref{Fgcox}) has two zeros:
one in
$k_{1\mathrm R}=\sqrt{\kappa_2/\kappa_1} \beta - \imath \alpha_1$,
$k_{2\mathrm R}=- \sqrt{\kappa_1/\kappa_2} \beta - \imath \alpha_2$,
the other one in $-k^*_{1\mathrm R}, -k^*_{2\mathrm R}$.
Since these zeros lie in the lower-half $k_1$ and in the upper-half $k_2$ complex planes,
they correspond to a resonance in channel 1,
only visible below threshold $\Delta=\kappa_2^2-\kappa_1^2$.
The three remaining parameters $\kappa_1$, $\kappa_2$ and $\beta$ can be
expressed in terms of this resonance energy $E_{\mathrm R}$ and width $\Gamma$,
as defined by $k^2_{1\mathrm R} \equiv E_{\mathrm R}-\imath \Gamma/2$.
They read
\beq
2\kappa_{1,2}^2 & = & \sqrt{E_{\mathrm R}^2+\Gamma^2/4}+
\sqrt{(E_{\mathrm R}-\Delta)^2+\Gamma^2/4} \mp \Delta, \eol
4\beta^4 & = & \left( E_{\mathrm R} + \sqrt{E_{\mathrm R}^2+\Gamma^2/4}\right) \eol
&& \times \left(E_{\mathrm R}-\Delta+\sqrt{(E_{\mathrm R}-\Delta)^2+\Gamma^2/4} \right).
\eeqn{physparam}

\begin{figure}
\scalebox{0.45}{\includegraphics{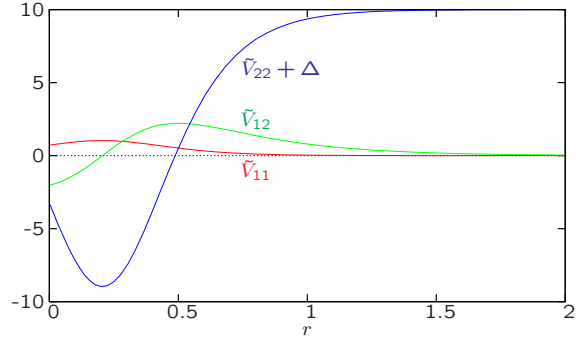}}
\caption{\label{fig:pot_gcox}
(Color online) Two-channel potential matrix (\ref{Vgcox}) as defined by parameters (\ref{physparam})
for threshold energy $\Delta=10$, 
Feshbach-resonance energy $E_{\mathrm R}=7$ and width $\Gamma=1$.}
\end{figure}

\begin{figure}
\scalebox{0.45}{\includegraphics{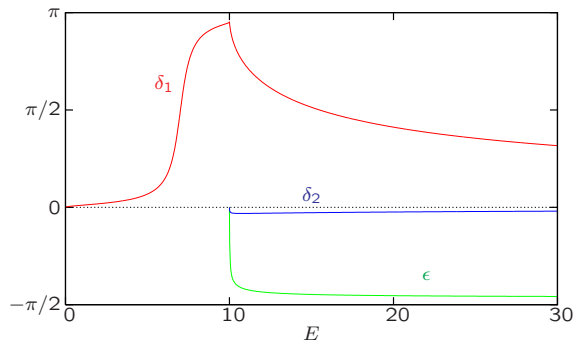}}
\caption{\label{fig:sanal_gcox}
(Color online) Eigenphases $\delta_1$, $\delta_2$ and mixing parameter $\epsilon$ of the
scattering matrix corresponding to the potential of Fig.~\ref{fig:pot_gcox},
as defined by Eqs.\ (\ref{S}), (\ref{Fgcox}), (\ref{alpha12}) and (\ref{physparam}).}
\end{figure}

Let us stress that, in general,
it is not trivial to calculate the zeros of the determinant of the
Jost matrices (\ref{Fcox}) and (\ref{Fgcox}),
i.e., to calculate the location of their bound states and/or resonances.
In the spirit of the inverse problem,
it is however very useful to construct potential models with bound
and resonant states corresponding to physical values.
The above potential provides such a model: it is a simple, exactly-solvable,
two-channel interaction with one controllable resonance.
When $E_{\mathrm R}<\Delta$,
this resonance is a Feshbach resonance,
for which the model constitutes a good pedagogical example:
when $\Gamma=0$,
all elements of $\tilde{V}$ vanish,
except for $\tilde{V}_{22}$ which has then a bound state at energy $E_{\mathrm R}$.
This corresponds to a zero of the Jost-matrix determinant
in $k_{1\mathrm R}=\sqrt{E_{\mathrm R}}$,
$k_{2\mathrm R}=\imath \sqrt{\Delta-E_{\mathrm R}}$.
When $\Gamma>0$,
coupling turns this bound state into a resonance by moving the zero off the axes,
counter-clockwise for $k_2$ and clockwise for $k_1$.
For instance, for $\Delta=10$, $E_{\mathrm R}=7$ and $\Gamma=1$,
one gets the potential plotted in Fig.\ \ref{fig:pot_gcox}.
The corresponding scattering matrix,
as defined by its eigenphases and mixing parameter \cite{newton:82,hesse:98},
is represented in Fig.~\ref{fig:sanal_gcox},
where the Feshbach resonance can be seen on $\delta_1$.
Crossing threshold,
which corresponds to going from positive imaginary $k_2$ to positive real $k_2$,
produces an interesting cusp effect in $\delta_1$ \cite{newton:82},
for which the present model also constitutes an analytical example.
The case $E_{\mathrm R}>\Delta$,
though allowed,
is less interesting from the physical point of view:
when $\Gamma=0$, coupling does not disappear and
the zero of the Jost-matrix determinant lies in
$k_{1\mathrm R}=\sqrt{E_{\mathrm R}}$,
$k_{2\mathrm R}=-\sqrt{E_{\mathrm R}-\Delta}$,
with no strong impact on the physical region $k_1>\sqrt{\Delta}, k_2>0$.

In conclusion, we have introduced a new type of supersymmetric transformations
which are able to transform an uncoupled potential into a potential with a non-diagonal
scattering matrix for non-vanishing thresholds 
(see the non-trivial behavior of $\epsilon$ in Fig.\ \ref{fig:sanal_gcox}).
This is a promising first step for coupled-channel inversion in the presence of thresholds.
The simplest possible application of this formalism leads to an exactly-solvable potential
which, for particular choices of parameters,
provides a textbook example of the Feshbach-resonance phenomenon,
with compact analytical expression both for the potential and its Jost matrix.
Future research will try to extend this result to iteration of transformations,
with the hope to get a supersymmetric-quantum-mechanics inversion technique
as efficient in the coupled-channel case as it is in the single-channel case \cite{baye:04}.

\acknowledgments
This text presents research results of the Belgian program P5/07 on interuniversity
attraction poles of the Belgian Federal Science Policy Office.
B.F.S.\ is partially supported by RFBR grant 06-02-16719 and
thanks the National Fund for Scientific Research, Belgium, for support
during his stay in Brussels.

\end{document}